\newcommand{\blind}{1}

\documentclass[12pt]{article}

\usepackage{amsmath,epsfig,amssymb,bm,amsfonts,amsthm,epsfig,verbatim, enumitem}
\usepackage[comma,authoryear]{natbib}
\usepackage{color,graphicx,lscape,longtable,mathabx,natbib, float, multirow}
\usepackage{hyperref,subfigure, multirow, setspace,  booktabs}
\usepackage{algorithm}
\usepackage{algorithmicx}
\usepackage{algpseudocode}
\graphicspath{{figures/}}

\renewcommand{\baselinestretch} {1.6}
\makeatletter 
\def\singlespace{\def\baselinestretch{1}\@normalsize}

\date{}
\newcommand{\bit}{\begin{itemize}}
\newcommand{\eit}{\end{itemize}}

\newcommand{\normmm}[1]{{\left\vert\kern-0.25ex\left\vert\kern-0.25ex\left\vert #1
    \right\vert\kern-0.25ex\right\vert\kern-0.25ex\right\vert}}

\@addtoreset{equation}{section}
\newtheorem{Theorem}{Theorem}[]

\renewcommand{\hat}{\widehat}
\def\bse{\begin{eqnarray*}}
\def\ese{\end{eqnarray*}}
\def\be{\begin{eqnarray}}
\def\ee{\end{eqnarray}}
\def\bsq{\begin{equation*}}
\def\esq{\end{equation*}}
\def\bq{\begin{equation}}
\def\eq{\end{equation}}

\def\wh{\widehat}

\def\bLambda{\boldsymbol\Lambda}

\def\beps{\boldsymbol\epsilon}

\def\bb{{\boldsymbol\beta}}

\def\bphi{\boldsymbol\phi}

\def\Delta{\boldsymbol\Delta}

\def\btheta{\boldsymbol\theta}

\newcommand{\nvs}{\vspace{-0.3in}}

\def\bbo{{\bf 0}}

\def\A{{\bf A}}
\def\U{{\bf U}}
\def\V{{\bf V}}

\def\bo{{\bf o}}

\def\a{{\bf a}}

\def\B{{\bf B}}

\def\D{{\bf D}}
\def\V{{\bf V}}
\def\K{{\bf K}}

\def\r{{\bf r}}
\def\f{{\bf f}}
\def\h{{\bf h}}
\def\b{{\bf b}}
\def\I{{\bf I}}

\def\Q{{\bf Q}}
\def\F{{\bf F}}
\def\H{{\bf H}}
\def\m{{\bf m}}

\def\K{{\bf K}}

\def\U{{\bf U}}
\def\S{{\bf S}}

\def\m{{\bf m}}
\def\v{{\bf v}}

\def\X{{\bf X}}
\def\S{{\bf S}}
\def\x{{\bf x}}
\def\I{{\bf I}}

\def\Y{{\bf Y}}

\def\y{{\bf y}}
\def\0{{\bf 0}}
\def\Z{{\bf Z}}
\def\z{{\bf z}}

\def\bSig{{\bf \Sigma}}

\def\bPhi{{\bf \Phi}}

\def\bq{\begin{equation}}
\def\eq{\end{equation}}

\def\wh{\widehat}

\def\trans{^{\rm T}}

\def\squarebox#1{\hbox to #1{\hfill\vbox to #1{\vfill}}}
\def\btheta{{\boldsymbol \theta}}
\def\1{{\bf 1}}

\def\bse{\begin{eqnarray*}}
\def\ese{\end{eqnarray*}}
\def\be{\begin{eqnarray}}
\def\ee{\end{eqnarray}}
\def\bsq{\begin{equation*}}
\def\esq{\end{equation*}}
\def\bq{\begin{equation}}
\def\eq{\end{equation}}
\def\wh{\widehat}

\def\trans{{\mathrm{T}}}
\def\boxit#1{\vbox{\hrule\hbox{\vrule\kern6pt\vbox{\kern6pt#1\kern6pt}\kern6pt\vrule}\hrule}}

\def\boxit#1{\vbox{\hrule\hbox{\vrule\kern6pt
          \vbox{\kern6pt#1\kern6pt}\kern6pt\vrule}\hrule}}

\catcode`@=11
\def\@evenhead{\vbox{\hbox to\textwidth{\tiny \hfill \hfill \today } }}
\def\@oddhead{\vbox{\hbox to \textwidth{\tiny \hfill \hfill \today } }}

\begin{document}

\def\spacingset#1{\renewcommand{\baselinestretch}%
{#1}\small\normalsize} \spacingset{1}
{
\renewcommand{\baselinestretch} {1}
\title{  \textbf{{High-Dimensional Multi-Study Multi-Modality Covariate-Augmented Generalized Factor Model
}}}

\if1\blind
\author{Wei Liu$^1$, Qingzhi Zhong$^{2*}$ \\
{\small$^1$School of Mathematics, Sichuan University, Chengdu, China}\\
{\small $^2$School of Economics, Jinan University, Guangzhou, China}
}
\fi

\maketitle
\thispagestyle{empty}

\if1\blind
\begin{singlespace}
\begin{footnotetext}
{*Corresponding author.   Email: \emph{zhongqz19@icloud.com}.}
\end{footnotetext}
\end{singlespace}
\fi

\bigskip
\begin{abstract}
Latent factor models that integrate data from multiple sources/studies or modalities have garnered considerable attention across various disciplines. However, existing methods predominantly focus either on multi-study integration or multi-modality integration, rendering them insufficient for analyzing the diverse modalities measured across multiple studies. To address this limitation and cater to practical needs, we introduce a high-dimensional generalized factor model that seamlessly integrates  multi-modality data from multiple studies, while also accommodating additional covariates.  We conduct a thorough investigation of the identifiability conditions to enhance the model's interpretability.  To tackle the complexity of high-dimensional nonlinear integration caused by  four large latent random matrices, we utilize a variational lower bound to approximate the observed log-likelihood by  employing a variational posterior distribution. By profiling the variational parameters, we establish the asymptotical properties of estimators for model parameters using M-estimation theory. Furthermore, we devise a computationally efficient variational EM algorithm to execute the estimation process and a criterion to determine the optimal number of both study-shared and study-specific factors. Extensive simulation studies and a real-world  application show that the proposed method significantly outperforms existing methods in terms of estimation accuracy and computational efficiency. The R package for the proposed method is publicly accessible at \url{https://CRAN.R-project.org/package=MMGFM}.
\end{abstract}

\noindent%
{\it Keywords:} Generalized factor model, Multiple modalities, Multiple studies, M-estimation,  Variational inference.
\vfill

\newpage
\spacingset{1.9} 

\section{Introduction}

Data derived from multiple sources/studies and modalities is increasingly prevalent across diverse fields. In education and psychology, researchers gather test-takers' responses to various test items tailored for distinct population groups, aiming to evaluate both the cognitive abilities within a group and the disparities across groups \citep{stenlund2018successful}.
Moreover, in the field of biology, the single-cell multi-omics sequencing technologies~\citep{vandereyken2023methods} enable the simultaneous measurement of multiple types of molecules across thousands of cells from diverse subjects or conditions, providing unprecedented insights into biological processes \citep{liu2023probabilistic}.
Given the significance of multi-study and multi-modality data in these and other domains, the integration of such data has become a crucial aspect of modern research and analysis.  Latent factor models have showcased a plethora of successful applications in both multi-study integration~\citep{argelaguet2018multi, de2019multi,avalos2022heterogeneous,chandra2024inferring} and multi-modality integration~\citep{welch2019single, li2022integrative, GFMLiu}.


In the context of multi-study integration, a pivotal aspect involves preserving the shared commonalities across studies while also discerning the unique variations specific to each, given their inherent heterogeneity. 
Therefore, \cite{liu2023probabilistic} and \cite{argelaguet2018multi} devised the multi-study linear factor analysis (MSFA) model, incorporating an additional study-unique factor term, enabling the extraction of both study-shared and study-specific features. 
In scenarios where additional covariate information is available, \cite{de2019multi} extended the MSFA framework by incorporating a regression term, yielding the multi-study factor regression (MSFR) model. However, MSFR has limitations in handling count-type data. Consequently, \cite{liu2024highdimensional} proposed covariate-augmented Poisson factor models for  multiple studies to better accommodate such data types.

For multi-modality integration, \cite{welch2019single} devised an innovative matrix factorization approach and \cite{argelaguet2018multi} introduced MOFA that links each modality matrix to a shared factor matrix and a modality-specific loading matrix. However, \cite{li2022integrative} adopted a different strategy, first applying linear factor analysis to  modality matrix individually and then combining  latent factors obtained from all modalities. Notably, both of these approaches overlook the variable type information embedded within the modality matrices and have limitations in capturing nonlinear latent factors. To address this issue, \cite{GFMLiu} proposed a tailored nonlinear latent factor model specifically designed for multiple modality matrices with diverse variable types. Despite these advancements, research on simultaneous multi-study and multi-modality integration remains scarce and needs for further exploration.

To address the practical need and methodological void, we have devised a multi-study, multi-modality, covariate-augmented generalized factor model, abbreviated as MMGFM. To the best of our knowledge, this model is being introduced for the first time in this work. We also theoretically examine the identifiability issue within this framework. Leveraging variational approximation, we transform the intricate model parameter estimation in the observed log-likelihood into a more manageable variational lower bound, encompassing both model and variational parameters. We also present a computationally efficient variational expectation-maximization algorithm to effectively carry out the estimation scheme. we derive the convergence rates of the estimators by bridging variational likelihood and profiled M-estimation. We develop a criterion, rooted in the step-wise singular value ratio of loading matrices, to ascertain the optimal number of both study-shared and study-specific factors. Finally, through numerical experiments, we demonstrate that MMGFM significantly surpasses existing methods in terms of estimation accuracy and computational efficiency.


The subsequent sections of the paper are structured as follows. In Section \ref{sec:model}, we offer an overview of the model setup and estimation method for MMGFM. Section \ref{sec:asymp} is dedicated to establishing the convergence rate and asymptotic normality of estimators. Following that, we detail the variational EM algorithm for MMGFM in Section \ref{sec:alg}, along with presenting the model selection method  in Section \ref{sec:modelselect}. To assess the performance of MMGFM, simulation studies are conducted in Section \ref{sec:simu}, and real data analysis is presented in Section \ref{sec:real}. In Section \ref{sec:dis}, we briefly discuss potential avenues for further research in this field. Technical proofs and additional materials are provided in the Supplementary Materials. 

\nvs
\section{Model and estimation} \label{sec:model}
\subsection{Proposed model}
Suppose we have $M$ modalities from $S$ distinct data sources. For each modality $m$ from source $s$, we observe a matrix $\X_{sm} = (x_{simj})_{i \leq n_s, j \leq p_m} \in \mathbb{R}^{n_s \times p_m}$, where $n_s$ is the number of observations and $p_m$ is the number of features for modality $m$. All variables in modality $m$ share the same type $t_m$. Additionally, we have covariates $\z_{si} \in \mathbb{R}^d$ providing contextual information for each unit.

To model the modality matrices with different variable types and from different sources,
we consider the framework of  generalized factor model~\citep{GFMLiu} augmented by covariates through a hierarchical formulation:
\begin{eqnarray}
\label{eq:xymodel}
&&  \hspace{-2em} x_{simj}|y_{simj} \stackrel{{c.i.d.}}\sim EF(g_m(y_{simj})),  s=1,\cdots, S, m=1, \cdots, M, \\
&& \hspace{-2em} y_{simj} = \tau_{sim} + \z_{si}^{\trans}\bb_{mj}+ \f_{si}^{\trans} \a_{mj}+ \h_{si}^{\trans}\b_{smj} + v_{sim} + \varepsilon_{simj},\label{eq:ymodel}
\end{eqnarray} 
for $i=1,\cdots, n_s,  j =1,\cdots, p_m, $
where ``$c.i.d.$" in model \eqref{eq:xymodel} implies that $\{x_{simj}, m=1,\ldots,M, j=1,\ldots, p_m\}$ is conditionally independent given $\{y_{simj}, m=1,\ldots,M, j=1,\ldots, p_m\}$, and {$EF(\cdot)$} is an exponential family distribution and $g_m(\cdot)$ is called the mean function for variable/modality type $t_m$.
For instance,  $EF(g_m(y_{simj}))$ is  $N(y_{simj}, 0)$ for a continuous type (i.e., $y_{simj}=x_{simj}$), $Poisson(\exp(y_{simj}))$ for a count type, and $Bernoulli(\frac{1}{1+\exp(-y_{simj})})$ for a binary type.
$\tau_{sim}$ is a known offset term for unit $i$ from source $s$ and modality $m$,  $\bb_{mj}$ is a modality-specified regression coefficient vector, $\f_{si}\in \mathbb{R}^q$ is the study-shared latent factor,
and $\a_{mj}$ is the corresponding loading vector, while $\h_{si} \in \mathbb{R}^{q_s}$ is the study-specified latent factor and $\b_{smj}$ is the corresponding loading vector, $v_{sim} $ is the study-specified and same modality variable-shared factor, and $\beps_{si}=(\varepsilon_{si11}, \cdots, \varepsilon_{si1p_1}, \cdots, \varepsilon_{siM1}, \cdots, \varepsilon_{siMp_M})^{\trans} \stackrel{i.i.d.}\sim N(\bbo, \bLambda_s)$ with $\bLambda_s= \mathrm{diag}(\lambda_{s11}, \cdots, \lambda_{s1p_1}, \cdots, $ $\lambda_{sM1}, \cdots, \lambda_{sMp_M})\in \mathbb{R}^{p \times p}$ and { $ p=\sum_{m=1}^{M} p_m$, where $\beps_{si}$ is able to capture the overdispersion of non-continuous variables~\citep{liu2024high}. }  We assume $\f_{si} \stackrel{i.i.d.}\sim N(\bbo, \I_q)$, $\h_{si} \stackrel{i.i.d.}\sim N(\bbo, \I_{q_s})$, and $v_{sim} \stackrel{i.i.d.}\sim N(0, \sigma^2_{sm})$, where $\I_q$ is a $q$-by-$q$ identity matrix.
{Models \eqref{eq:xymodel} and \eqref{eq:ymodel} are not identifiable, we establish their identifiability  and provide proofs in the proposition presented in Appendix A of the Supplementary Materials.}

\nvs
\subsection{Estimation}
Analogous to \cite{GFMLiu}, we primarily focus on three variable/modality types: continuous, count, and categorical, as they are widely encountered in practical applications. Thus, $t_m \in G$,  $G = \{1, 2, 3\}$, where each element represents a distinct type, namely, continuous, count, and categorical. { Additionally, the estimation procedure and corresponding algorithm for other types within the exponential family can be formulated in a similar manner; see Appendix B in Supplementary Materials for more discussion.}  The model \eqref{eq:xymodel} for these three modality types ($t_m \in G$) is explicitly expressed as:
\begin{eqnarray}
  && x_{simj} = y_{simj}, \ \mbox{if } t_m =1, \quad\quad x_{simj}|y_{simj} \sim Poisson(\exp(y_{simj})) , \ \mbox{if } t_m =2 , \label{eq:G3x} \\
  &&  P(x_{simj}=k|y_{simj}) = C_{n_{mj}}^k p_{simj}^k (1-p_{simj})^{n_{mj}-k},
  p_{simj}=\frac{1}{1+\exp(-y_{simj})}, \ \mbox{if } t_m =3, \nonumber
\end{eqnarray}
where $n_{mj}$ is the number of trials for the $j$-th variable in the modality $m$. If $n_{mj}=1$ for all $j$'s in the modality $m$, the binomial variable $x_{simj}$ reduces to the 0-1 variable with success
probability $p_{simj}$.

Let $\x_{si}=(x_{si11}, \cdots, x_{si1p_1}, \cdots, x_{siMp_M})^{\trans},\y_{si}=(y_{si11}, \cdots, y_{si1p_1}, \cdots, y_{siMp_M})^{\trans}, \v_{si}=(v_{si1}, $ $ \cdots,v_{siM})^{\trans},\bb_m=(\bb_{m1},\cdots,\bb_{mp_m})^{\trans},\A_m = (\a_{m1}, \cdots, \a_{mp_m})^{\trans}, \B_{sm} = (\b_{sm1}, \cdots, \b_{smp_m})^{\trans}$, { $ \btheta_s=(\mbox{tvec}\{\A_m,  \bb_m, \B_{sm}\},\mbox{tvec}\{\bLambda_s\}, \sigma^2_{sm}, m=1, \cdots,M)^{\trans}$ be a column vector stacked by $m$ and $\btheta= \{\btheta_s^{\trans}, s=1, \cdots, S\}^{\trans}$ be a column vector obtained by stacking $\btheta_s^{\trans}$ across $s$. Here, $\mbox{tvec}\{\bLambda_s\}$ denotes the vectorization
of $\bLambda_s$ followed by transposition. Note that $\btheta$ is}  the collection of  model parameters from all studies. Denote $\L_{si} = (\x_{si}^{\trans}, \z_{si}^{\trans})^{\trans}$ and $P(\x_{si}, \y_{si}, \f_{si}, \h_{si}, \v_{si}) =P(\x_{si}|\y_{si})P(\y_{si}| \z_{si}, \f_{si}, \h_{si}, \v_{si}) P(\f_{si})P(\h_{si})P(\v_{si})$, then the observed log-likelihood for the unit $i$ in study $s$ is given by
\begin{equation}\label{eq:loglike}
    l(\btheta_s; \L_{si}) = \ln \int P(\x_{si}, \y_{si}, \f_{si}, \h_{si}, \v_{si}) d\y_{si} d\f_{si}d\h_{si}d \v_{si}.
\end{equation}
Theoretically, to obtain the maximum likelihood estimator of $\btheta$, we are able to maximize the observed log-likelihood for all samples: $l(\btheta)=\sum_{s=1}^S \sum_{i=1}^{n_s} l(\btheta_s; \L_{si})$. However, the practical implementation is very challenging and intractable since the integrated function is a highly nonlinear and high-dimensional function with respect to four groups of integral variables. 

Variational approximation is a powerful tool for approximating distributions, efficiently transforming the challenging integration of latent variables into a more manageable optimization problem involving variational parameters. In the existing literature, numerous studies have harnessed a variational lower bound to approximate the observed log-likelihood~\citep{wang2019frequentist, pang2023factor}. Similarly, we adopt the same  strategy to handle \eqref{eq:loglike}. { Let $ \mathcal{G}_j=\{m: t_m =j\}, j\in G$, $\mathcal{G}=\mathcal{G}_2 \cup \mathcal{G}_3$ and $\y_{sim}=(y_{sim1}, \cdots, y_{simp_m})^{\trans}$. Then, we define all latent random variables: $\Y_{\mathcal{G}}=\{\y_{sim},m\in \mathcal{G}\}, \F=\{\f_{si}\}, \H=\{\h_{si}\}$ and $\V=\{\v_{si}\}$.} If $\mathcal{G}\neq \emptyset$, to manage the four large latent random matrices $(\Y_{\mathcal{G}}, \F, \H, \V)$ within the observed log-likelihood, we define
 \begin{equation}\label{eq:etaY}
    \eta(\Y_{\mathcal{G}}, \F, \H, \V) = \Pi_{s=1}^S\Pi_{i=1}^{n_s} \eta_{si}(\y_{si}, \f_{si}, \h_{si}, \v_{si}),
\end{equation}
with $\eta_{si}(\y_{si}, \f_{si}, \h_{si}, \v_{si})= \Pi_{m \in \mathcal{G}} \Pi_{j=1}^{p_m} \left\{N(y_{simj}; \xi_{simj}, \sigma^2_{simj})  N(\f_{si}; \m_{si}, \bSig_{si})  N(\h_{si}; \bo_{si}, \bPhi_{si})\right\}$ $ \times \Pi_{m=1}^M N(v_{sim}; w_{sim}, \varsigma_{sim})$, which belongs to a mean field varitional distribution family, where $N(y;\xi, \sigma^2)$ denotes a normal density function of $y$ with mean $\xi$ and variance $\sigma^2$. Let $\bphi_{si}=(\xi_{simj}, \sigma^2_{simj}, \m_{si}^{\trans}, \mbox{tvec}\{\bSig_{si}\}, \bo_{si}^{\trans}, \mbox{tvec}\{\bPhi_{si}\}, w_{sim}, \varsigma_{sim},  m=1, \cdots, M, j=1, \cdots, p_m)^{\trans}$, then $\bphi=(\bphi_{si}^{\trans},s=1,\cdots,S, i=1,\cdots, n_s)^{\trans}$ is the collection of variational parameters. {By minimizing the Kullback-Leibler (KL) divergence between $\eta_{si}(\y_{si}, \f_{si}, \h_{si}, \v_{si})$ and the posterior density of $(\y_{si}, \f_{si}, \h_{si}, \v_{si})$ with respect to $\bphi_{si}$, we identify the optimal variational density $\eta_{si}(\y_{si}, \f_{si}, \h_{si}, \v_{si})$ with variational parameters $\wh\bphi_{si}(\btheta_s)$,  which is closest in KL divergence to the posterior
density of $(\y_{si}, \f_{si}, \h_{si}, \v_{si})$ within the mean field varitional distribution family. We define the variational approximated  log-likelihood for the unit $i$ in study $s$ as
$$\tilde l(\btheta_s, \bphi_{si}; \L_{si}) = \int \ln \left(\frac{P(\x_{si}, \y_{si}, \f_{si}, \h_{si}, \v_{si}) }{\eta_{si}(\y_{si}, \f_{si}, \h_{si}, \v_{si})}\right) \eta_{si}(\y_{si}, \f_{si}, \h_{si}, \v_{si}) d\y_{si} d\f_{si}d\h_{si}d \v_{si}.$$
Therefore,  $\tilde l(\btheta_s, \wh\bphi_{si}(\btheta_s); \L_{si})$, where $\wh\bphi_{si}(\btheta_s)=\arg\max_{\bphi_{si}} \tilde l(\btheta_s, \bphi_{si}; \L_{si})$, provides a close approximation  to $l(\btheta_s; \L_{si})$, with the degree of closeness measured by KL divergence.
By Jensen's inequality, we obtain $l(\btheta_s; \L_{si}) \geq \tilde l(\btheta_s, \bphi_{si}; \L_{si})$ and $l(\btheta)\geq \tilde l(\btheta, \bphi) = \sum_{s=1}^S \sum_{i=1}^{n_s}$ $ \tilde l(\btheta_s, \bphi_{si}; \L_{si})$,  with equality if and only if \eqref{eq:etaY} is equal to the posterior density of $(\Y_{\mathcal{G}}, \F, \H, \V)$. This implies that there exists $\bphi$ such that $l(\btheta) = \tilde l(\btheta, \bphi)$ if the posterior density of $(\Y_{\mathcal{G}}, \F, \H, \V)$ belongs to the presumed variational distribution family. Here, $\tilde l(\btheta, \bphi)$ is referred to as the {\it evidence lower bound} or {\it variational lower bound} of $l(\btheta)$~\citep{wang2019frequentist}. We derive the estimator of $\btheta$ as $\wh\btheta$, where $\wh\btheta$ are obtained by maximizing $\tilde l(\btheta, \bphi)$ over $(\btheta,\bphi)$. We refer to $\wh\btheta$ as the maximum variational lower bound estimator.  }

\nvs
\section{Asymptotical analysis}\label{sec:asymp}
We establish the large-sample properties for $\wh\btheta$ by employing M-estimation theory on the profiled version of $\tilde l(\btheta, \bphi)$, with detailed proofs provided in Appendix C of the Supplementary Materials. 
Prior to presenting our main findings, we provide some essential definitions and notations.
{ Let $D_\bullet,D^2_\bullet,\lambda_{\min}(\bullet)$ and $\|\bullet\|_{\max}$ denote the first, second derivative, smallest eigenvalue and  the maximum absolute value of the elements} with respect to $\bullet$, respectively, and $\|\bullet\|$ denotes either the Euclidean norm for a vector or the Frobenius matrix norm for a matrix. For a column vector $\a$, we define $\mathrm{diagmat}(\a)$ as an operator that transforms $\a$ into a diagonal matrix.

Firstly, we establish a connection between maximum variational lower bound estimation and M-estimation.
Let $\wh\bphi(\btheta)=(\wh\bphi_{si}(\btheta_s)^{\trans},s=1,\cdots,S,i=1,\cdots, n_s)^{\trans}$. 
Then,  the profiled version $\tilde l_p(\btheta)=\sum_{s,i} \tilde l_p(\btheta_s;\L_{si})=\sum_{s,i}\tilde l(\btheta_s, \wh\bphi_{si}(\btheta_s);\L_{si})$ is called {\it variational log-likelihood}, and the {\it maximum variational log-likelihood estimator} of $\btheta$ is given by $\wh\btheta^*=\arg\max \tilde l_p(\btheta)$, posing it as an M-estimation problem. Then, we have following results.
\nvs
\begin{Theorem}
 (i) For all $\btheta_s$, $\wh\bphi_{si}(\btheta_s)$ is the unique maximizer of $\tilde l(\btheta_s, \bphi_{si}; \L_{si})$; (ii) the maximum varitional lower bound estimator is equal to the  maximum variational log-likelihood estimator, i.e., $\wh\btheta=\wh\btheta^*$.
\end{Theorem}
\nvs
Subsequently, we impose the following conditions to derive convergence rates and establish asymptotic normality.
\begin{description}
\item [(C1)] Conditions (A1)--(A4) in Appendix A of the Supplementary Materials  hold.
\item [(C2)] There exists some positive constant $M$ such that $\|\btheta\|_{\max}\leq M$ and $\mathrm{E}\|\z_{si}\|^2\leq M$.
\item [(C3)]  There exists positive constant $c$ such that $\lambda_{\min}\{\mathrm{diagmat}(\r_n)\K_1\mathrm{diagmat}(\r_n)\}>c$, where
{ $\r_n=(\1^{\trans}_{p_mq+p_md}/\sqrt{\sum_{s}n_s},\1^{\trans}_{p_mq_s+p_m+1}/\sqrt{n_s},m\leq M,s\leq S)^{\trans}$ is a column vector fisrt stacked by $m$ and then by $s$ and $\K_1= \mathrm{E}\{-D^2_{\btheta\btheta}\sum_{s,i} \tilde l_p(\btheta_{s0};\L_{si})\}$.}
\end{description}
Condition (C1) guarantees the identifiability of the model. Conditions (C2) and (C3) are frequently employed in prior research on variational approximation; see \cite{liu2024highdimensional} and \cite{pang2023factor}. Recalling $\btheta= (\btheta_s^{\trans}, s=1, \cdots, S)^{\trans}$ with $ \btheta_s=(\mbox{tvec}\{\A_m,  \bb_m, \B_{sm}\},$\\ $\mbox{tvec}\{\bLambda_s\},  \sigma^2_{sm}, m=1, \cdots,M)^{\trans}$.
\begin{Theorem}\label{thm3}
Let $\btheta_0=\arg\max_{\btheta}\mathrm{E}\{\sum_{s,i} \tilde l_p(\btheta_{s};\L_{si})\}$.
If $p_m=O\{(\sum_s n_s)^{\rho}\}$ for any given $m$, $(\sum_s n_s)^{4\rho}=o(\min_sn_s)$ and Conditions (C1)-(C3) hold, then
\begin{eqnarray*}
&&\|\widehat\bb_m-\bb_{m0}\|^2+\|\widehat\A_m-\A_{m0}\|^2=O_p(\frac{p_m}{\sum_sn_s}),\\
&& \|\widehat\B_{sm}-\B_{sm0}\|^2+\|\widehat\bLambda_{sm}-\bLambda_{sm0}\|^2+|\widehat\sigma^2_{sm}-\sigma^2_{sm0}|=O_p(\frac{p_m}{n_s}).
 \end{eqnarray*}
\end{Theorem}

Theorem \ref{thm3} demonstrates that the estimated study-shared and modality-specific parameters converge at a rate of $O_p(\sqrt{\frac{p_m}{\sum_sn_s}})$, while the estimated study-specific and modality-specific parameters converge at a rate of $O_p(\sqrt{\frac{p_m}{n_s}})$. This difference in convergence rates arises from the fact that estimating study-shared and modality-specific parameters utilizes information from all $\sum_s n_s$ individuals, whereas estimating study-specific and modality-specific parameters relies solely on information within a single study.

Next, we derive the asymptotical expression for all model parameters, and then establish asymptotical normality for estimator at the single  study and modality level.
\begin{Theorem}\label{thm4}
{ Define $o_p(\r_n)$ as the application of $o_p(\bullet)$ to each element of $\r_n$.} Under the conditions in Theorem \ref{thm3}, it holds that
$$\widehat\btheta-\btheta_0=\K_1^{-1}D_{\btheta}\sum_{s,i}\tilde l_p(\btheta_{s0};\L_{si})+o_p(\r_n).$$
Let $\Q=\{\mathrm{diagmat}(1/\r_n)\K_1^{-1}\K_2\K_1^{-1}\mathrm{diagmat}(1/\r_n)\}$, where $\K_2=\mathrm{E}\{D_{\btheta}\sum_{s,i}\tilde l_p(\btheta_{s0};\L_{si})$ \\ $D^{\trans}_{\btheta}\sum_{s,i}\tilde l_p(\btheta_{s0};\L_{si})\}$. Denote $\Q_{\bb_{mj}\bb_{mj}},\Q_{\a_{mj}\a_{mj}},\Q_{\b_{smj}\b_{smj}},\Q_{\lambda_{smj}\lambda_{smj}}$ and $\Q_{\sigma^2_{sm}\sigma^2_{sm}}$ are submatrices of $\Q$ corresponding to $\bb_{mj},\a_{mj},\b_{smj},\lambda_{smj}$ and $\sigma^2_{sm}$, respectively.  Then, it holds that for modality-specified parameters,
 \begin{eqnarray*}
&& \sqrt{\sum_s n_s}(\widehat\bb_{mj}-\bb_{mj,0})\rightarrow N(\0,\Q_{\bb_{mj}\bb_{mj}}),\\
&&\sqrt{\sum_s n_s}(\widehat\a_{mj}-\a_{mj,0})\rightarrow N(\0,\Q_{\a_{mj}\a_{mj}}),
  \end{eqnarray*}
  and for study- and modality-specified parameters, we have
   \begin{eqnarray*}
&&\sqrt{n_s}(\widehat\b_{smj}-\b_{smj,0})\rightarrow N(\0,\Q_{\b_{smj}\b_{smj}}),\\
&&\sqrt{n_s}(\widehat\lambda_{smj}-\lambda_{smj,0})\rightarrow N(0,\Q_{\lambda_{smj}\lambda_{smj}}),\\
&&\sqrt{n_s}(\widehat\sigma^2_{sm}-\sigma^2_{sm,0})\rightarrow N(0,\Q_{\sigma^2_{sm}\sigma^2_{sm}}).
  \end{eqnarray*}
\end{Theorem}
The results outlined in Theorem \ref{thm4} broaden the scope of the discoveries made in \cite{liu2024highdimensional} to a covariate-augmented factor model in a multi-study, multi-modality framework. Within this context, we rigorously demonstrate the asymptotic normality of the estimator for model parameters that are associated with a wide range of modality types and data sources.
\nvs
\section{Implementation}
To implement MMGFM, we devise a variational EM algorithm and introduce a criterion for determining the optimal number of both study-shared and study-specific factors.

\subsection{Algorithm}\label{sec:alg}
The variational EM algorithm iteratively updates the variational parameters in the E-step and the model parameters in the M-step, respectively, in an alternating fashion. However, it is difficult to evaluate the variational parameters $(\xi_{simj}, \sigma^2_{simj})$  because $\eta_y(\Y_{\mathcal{G}})=:\Pi_{s,i}\Pi_{m \in \mathcal{G}} \Pi_{j=1}^{p_m}N(y_{simj};$ $\xi_{simj}, \sigma^2_{simj})$ is not a conjugate distribution to $P(\X|\Y_{\mathcal{G}})$, where $\X=\{\x_{si}\}$. We turn to the Laplace approximation combined with Taylor approximation  to update these variational parameters. Subsequently, we update the remaining parameters using explicit iterative solutions derived from the variational lower bound function $\tilde l(\btheta, \bphi)$. To conserve space, the comprehensive details of the algorithm, including pseudo-code and convergence properties, are provided in Appendix B of the Supplementary Materials.
\nvs
\subsection{Model selection}\label{sec:modelselect}
 The implementation of MMGFM necessitates identifying the numbers of study-shared ($q$) and study-specific ($q_s$) factors. Within the context of multi-study multi-modality data, the study-shared factor part, which aggregates information from all data sources, generally  exhibits a stronger signal compared to the study-specific factor parts. To account for this characteristic, we propose a step-wise singular value ratio (SVR) method.
Specifically, we fit our model with upper bounds on the number of factors, $q=q_{\max}$ and $q_s=q_{s,\max}$, and denote the estimated loading matrices as $\wh\A_m^{(\max)}$. We then estimate the numbers of study-shared factors for each modality by $\wh q_{\A_m}=\arg\max_{k \leq q_{\max}-1} \frac{\vartheta_k(\wh\A_m^{(\max)})}{\vartheta_{k+1}(\wh\A_m^{(\max)})}$, where $\vartheta_k(\wh\A_m^{(\max)})$ is the $k$-th largest singular value of $\wh\A_m^{(\max)}$. If the values of $\wh q_{\A_m}$  differ across modalities, we define the estimator for $q$ as $\wh q=\arg\max_{m}\wh q_{\A_m}$; otherwise, $\wh q$ is taken as the mode of the set $\{\wh q_{\A_m},m\leq M\}$. Subsequently, we refit our model with  the selected $\wh q$ and upper bounds on the number of study-specific factors $q_s=q_{s,\max}$. We then calculate the singular value ratio for the estimated loadings $\wh\B_{sm}^{(\max)}$ to estimate the numbers of study-specific  factors for each modality, applying the same principle to determine $\wh q_s$.

\nvs
\section{Simulation study}\label{sec:simu}
In simulation studies, we demonstrate the effectiveness of the proposed MMGFM through 100 replications, comparing it against several state-of-the-art methods: the generalized factor model~\citep{GFMLiu} (R package {\bf GFM}), the mixed-outcome reduced-rank regression (MRRR) model~\citep{luo2018leveraging} ({\bf rrpack} R package), the multi-study linear factor regression (MSFR) model~\citep{de2023multi} (available at \url{https://github.com/rdevito/MSFR}), and the multi-study covariate-augmented overdispersed Poisson factor (MultiCOAP) model~\citep{liu2024highdimensional} ({\bf MultiCOAP} R package).

We are interested in the estimation of six quantities: $\F, \H, \V, { \bb=(\bb_1^{\trans}, \cdots, \bb_M^{\trans})^{\trans} \in \mathbb{R}^{p \times d}}$, $\A=(\A_m)$, and $\B=(\B_{sm})$. Here, $\F$, $\H$, and $\V$ represent low-dimensional latent factor matrices for multi-modality, multi-study data, capturing study-shared, study-specific, and modality-shared information, respectively. $\bb_m$ quantifies covariate effects on modality $m$ variables, $\A_m$ measures variable contributions to study-shared factors, and $\B_{sm}$ assesses contributions to study-specific factors.

To measure the performance of $\F,\H, \V, \A$ and $\B$, we utilize the trace statistics, defined as $\mathrm{Tr}(\wh\D,\D_0)=\frac{\mathrm{Tr}\{\D_0^{\trans} \wh\D(\wh\D^{\trans}\wh\D)^{-1}\wh\D^{\trans}\D_0\}}{\mathrm{Tr}(\D_0^{\trans}\D_0)}$, as used in \cite{liu2024high}, which measures the similarity of two matrices. A higher value means better performance. Specifically, for $\F$ and $\B$, we calculate the mean of trace statistics, i.e., $MT_{\F}=\frac{1}{S}\sum_{s=1}^S \mathrm{Tr}(\wh\F_s,\F_{s0}), MT_{\B}=\frac{1}{S M}\sum_{s=1}^S\sum_{m=1}^M \mathrm{Tr}(\wh\B_{sm}, \B_{sm0})$, then $MT_{\H}, MT_{\V}$ and $MT_{\A}$ are defined similarly. To quantify the performance of regression coefficient matrix, we adopt the mean absolute error, defined as  $ME_{\bb}=\frac{\sum_{j=1}^p\sum_{k=1}^d|\beta_{jk} - \beta_{jk0}|}{d p}$.

\noindent\underline{\bf Scenario 1}: To compare with MultiCOAP, we consider three Poisson-distributed modalities ($M=3$) from  three data sources ($S=3$). Specifically, we set sample sizes $(n_1, n_2, n_3)=(300, 200, 100)$, variable dimensions $(p_1, p_2, p_3) = (50, 150, 200)$, covariate dimension $d=3$, common factors $q=3$, and study-specific factors $q_s=2$. Then, we generate $\bb_m$ with i.i.d. $N(0, 4)$ entries, $\breve{\B}_{1m} \in \mathbb{R}^{p1\times (q+q_1)}$ with i.i.d. $N(0,1)$ entries. To ensure identifiability, we apply SVD: $\breve{\B}_{1m}=\U_{1m} \S_{1m} \V_{1m}^{\trans}$, and let $\breve{\B}_{1m0}= \rho_m \U_{1m} \S_{1m}$, then define $\A_{m0}$ be the first $q$ columns of  $\breve{\B}_{1m0}$ and $\B_{1m0}$ be the last $q_1$ columns of $\breve{\B}_{1m0}$, where $\rho_m=2$ controls the signal strength of each modality. For $s>1$, we generate $\breve{\B}_{sm} \in \mathbb{R}^{p_m\times q_s}$  and define $\B_{sm0} \in \mathbb{R}^{p_m\times q_s}$ similarly. We generate $\Z_s = (\bm{1}_{n_s}, \tilde\Z_s)$ with $\tilde z_{sij}=\rho_z e_{sij}$ and $e_{sij} \stackrel{i.i.d.}\sim  U[-3, 3]$, $\f_{si}  \stackrel{i.i.d.}\sim N(\bbo,\I_q)$, $\h_{si}  \stackrel{i.i.d.}\sim N(\bbo,\I_{q_s})$ and $\varepsilon_{simj} \stackrel{i.i.d.}\sim N(0,1)$, where $\rho_z=0.5$ controls the signal strength of the covariates.  After generating $\bb_{m0}, \A_{m0}$ and $ \B_{sm0}$, they are fixed in repetition. Finally, we simulate $v_{sim} \sim N(0,\sigma_{sm}^2)$ with $\sigma_{sm}^2 \in \{0, 1, 3\}$ to explore correlation effects. Without loss of generality, we set the offset term $\tau_{sim}=0$. Then, by model \eqref{eq:ymodel}, we know $y_{simj} = \z_{si}^{\trans}\bb_{mj0}+ \f_{si}^{\trans} \a_{mj0}+ \h_{si}^{\trans}\b_{smj0} + v_{sim} + \varepsilon_{simj}$. In this Scenario, $x_{simj} = Poisson(y_{simj})$.

Table \ref{Scen_1} suggests that MMGFM outperforms the four compared methods in the estimation of factors and regression coefficient matrix in all runs, while performs worse than MultiCOAP in the estimation of loadings when $\sigma_{sm}^2=0$ and maintains a substantial lead when $\sigma_{sm}^2>0$. The lower performance of MMGFM in the $\sigma_{sm}^2=0$ case is due to the estimation cost of zero-valued $v_{sim}$ affecting the estimation of loadings. Notably, GFM  lacks the capability  to account for both covariates and multiple studies; MRRR treats the  coefficient matrix and factor part as a single entity  and is unable to account for multiple studies; MSFR accommodates the count nature of variables through log transformation on the count value of $x_{simj}$, while this preprocessing may not accurately capture the underlying data generation mechanism. Moreover, all the compared methods disregard the correlation within a modality, which is captured by $\V$, leading to subpar results. Consequently, MRRR consistently performs the worst across all values of $\sigma_{sm}^2$, while the performance of GFM, MSFR, and MultiCOAP declines significantly as $\sigma_{sm}^2$ increases.
\begin{table}[htbp]
  \centering
  \caption{Comparison of MMGFM  and other methods for parameter estimation.  Reported are the average (standard deviation) for performance metrics in Scenarios 1 and 2. Note $\V$ does not exist in the setting of Scenario 2 due to $\sigma^2_{sm}=0$, thus, $MT_{\V}$ is not available for MMGFM.}
    \resizebox{\textwidth}{!}{
    \begin{tabular}{rrrrrrrr}
      \hline
     Method &$\sigma_{sm}^2$ &$MT_{\F}$ &$MT_{\H}$ &$MT_{\V}$ &$MT_{\A}$ &$MT_{\B}$ &$ME_{\bb}$\\
     \hline
      &\multicolumn{7}{c}{Scenario 1}\\
      \cmidrule(lr){2-8}
   MMGFM&0   &0.97(4E-3) &0.98(1E-3) &-(-)      &0.72(0.03)  &0.76(0.02)  &0.29(0.01)  \\
          &1  &0.77(0.06)  &0.90(0.10)  &0.90(0.01)  &0.88(0.05)  &0.66(0.09)  &0.26(0.01)  \\
          &3     &0.54(0.07)  &0.63(0.11)  &0.82(0.01)  &0.69(0.04)  &0.43(0.08)  &0.28(0.01)  \\
           \hline
 GFM &0  &0.35(0.01)  &-(-)   &-(-)  &0.36(0.01)  &-(-) &-(-) \\
          &1  &0.06(0.02)  &-(-)   &-(-) &0.13(0.04)  &-(-) &-(-)\\
          &3     &0.02(0.01)  &-(-) &-(-) &0.10(0.03)  &-(-) &-(-)\\
           \hline
   MRRR &0  &0.02(0.01)  &-(-) &-(-)  &0.03(2E-3) &-(-)    &1.31(0.05)  \\
          &1  &0.02(0.01)  &-(-) &-(-)  &0.05(3E-3) &-(-)  &1.30(0.04)  \\
          &3   &0.02(0.01)  &-(-) &-(-) &0.05(2E-3) &-(-) &1.32(0.05)  \\
           \hline
  MSFR &0  &0.88(0.08)  &0.74(0.10) &-(-)      &0.77(0.07)  &0.58(0.08)  &1.09(4E-3) \\
          &1  &0.41(0.06)  &0.42(0.08)  &-(-) & 0.48(0.06)  &0.34(0.06)  &1.15(0.01)  \\
          &3   &0.19(0.10)  &0.30(0.08)  &-(-)  &0.36(0.07)  &0.24(0.06)  &1.20(0.01)  \\
           \hline
  MultiCOAP &0  &0.88(0.07)  &0.98(0.03)  &-(-)  &0.94(0.07)  &0.92(0.04)  &0.28(0.02)  \\
          &1  &0.44(0.07)  &0.46(0.08)  &-(-)  &0.62(0.07)  &0.41(0.07)  &0.57(0.09)  \\
          &3  &0.17(0.08)  &0.32(0.08)  &-(-)  &0.40(0.07)  &0.29(0.07)  &0.60(0.07)  \\
           \hline
            &\multicolumn{7}{c}{Scenario 2}\\
      \cmidrule(lr){2-8}
      MMGFM &&0.99(2E-3) &1.00(2E-4) &-(-)&0.98(2E-3) &0.93(3E-3) &0.30(5E-3)  \\
    GFM   &&0.14(0.12)  &-(-) &-(-)&0.23(0.06)  &-(-)    &-(-) \\
    MRRR  &&0.02(0.01) &-(-)  &-(-)&0.01(8E-4) &-(-) &1.43(0.03)  \\
    MSFR  &&0.98(3E-3) &1.00(3E-4) &-(-)&0.90(2E-3) &0.85(0.01)  &0.65(0.01)  \\
       \hline
    \end{tabular}%
    }
  \label{Scen_1}%
\end{table}%

\noindent\underline{\bf Scenario 2}.
In this Scenario, we explore the impact of different variable types. We consider three data sources and five modalities, i.e., $S=3, M=5$, with the first two modalities from Gaussian distributions and last three modalities from Poisson distributions.
We set $(p_1, p_2)=(100, 200), (p_3, p_4, p_5) = (50, 150, 200)$ and $\sigma_{sm}^2 = 0$, while maintaining other parameters consistent with Scenario 1. We compare MMGFM with GFM, MRRR and MSFR, as these
three methods can handle the mixed type of Gaussian and Poisson variables. The results in Table \ref{Scen_1} show that compared to the competing methods, MMGFM achieves the largest estimators of $MT_{\F},MT_{\H},MT_{\A},MT_{\B}$ and smallest estimator of $ME_{\bb}$. This indicates MMGFM's superior ability to account for different type of variables.

\noindent\underline{\bf Scenario 3}. Mixed types and intramodal correlations may coexist in observed data, prompting our investigation into their impact. Therefore, we consider six cases, emcompassing different combinations of modality types.

\noindent\underline{\bf Case 3.1}. Initially, we delve into the analysis of two data sources that encompass a blend of first three continuous modalities stemming from Gaussian distributions and last two binary modalities derived from  Bernoulli distributions.  We set  $(p_1, p_2,p_3)=(50, 150, 200), (p_4, p_5) = (100,60),\sigma_{sm}^2 = 0.7,(n_1, n_2)=(300, 200)$  and other parameters as in Scenario 1.

\noindent\underline{\bf Case 3.2}. We consider data generated such that the first three count modalities stem from Poisson distributions, while the last two binary modalities originate from Bernoulli distributions. We set $(p_1, p_2,p_3)=(50, 150, 200), (p_4, p_5) = (100,200)$, $\sigma_{sm}^2 = 0.5$ and other settings same as those in Case 3.1.

\noindent\underline{\bf Case 3.3}. Here, we delve into a more intricate case involving five modalities of three distinct types, where the first two modalities are from Gaussian distributions, third modality is from Poisson distribution and last two modalities are from Bernoulli distributions. We consider the influence of sample size by  varying  $(n_1, n_2)$  within $\{(50, 200), (300, 200), (300, 400)\}$, while fixing $(p_1, p_2)=(50, 150)$ and $ p_3=50, (p_4, p_5) = (100,60)$. The remaining settings are identical to those outlined in Case 3.2.

\noindent\underline{\bf Case 3.4}. On the basis of Case 3.3, we investigate the effect of variable dimension by varying $(p_1,p_2,p_3,p_4,p_5)\in\{(250,50,50,100,50),(250,50,250,100,50),(250,250,250,$ $100,50)\}$ while fixing $(n_1, n_2)=(100, 200)$. Other settings are same as those in Case 3.3.

\noindent\underline{\bf Case 3.5}. Based on Case 3.3, we continue considering the influence of the number of data sources by setting $S=20$ with $n_s=50$ for $s\leq 10$ and $n_s=30$ for $10<s\leq 20$.

\noindent\underline{\bf Case 3.6}. Finally, we consider the effect of the number of modalities. We set $M=20$,  with $p_m=150$ for $m\leq 7$, $p_m=50$ for $8<m\leq 13$ and $p_m=60$ for $13<m\leq 20$.

{ Table \ref{Scen_3} provides a summary of the average (with standard deviation) performance metrics, excluding MultiCOAP due to its design specifically for count variables and inability to handle variables with negative values.}  The findings demonstrate that MMGFM consistently surpasses GFM, MRRR, and MSFR across all modality types (Case 3.1 and 3.2), sample sizes (Case 3.3), variable dimensions (Case 3.4), and number of data sources (Case 3.5) and modalities (Case 3.6). Notably, the enhancement achieved by MMGFM is substantial in comparison to its competitors, as it is the only method that successfully accommodates multiple studies and modalities simultaneously, which the other methods fail to do.

Comparing the results in Cases 3.1 and 3.2, we observe that GFM and MRRR are very sensitive to the modality types. In Case 3.3, augmenting the sample size significantly enhances the estimation accuracy of MMGFM for both loadings and factors. This improvement is attributed to the fact that estimating $\a_{mj}$ and $\b_{smj}$ leverages information from the $j$th variable across $\sum_s n_s$ and $n_s$ individuals, respectively, and more precise loadings estimates positively influence the estimation of factors. In Case 3.4, elevating the variable dimension boosts the estimation performance of factors, loadings, and the regression coefficient matrix. This is because estimating $\f_{si}$ and $\h_{si}$ capitalizes on information from $\sum_m p_m$ variables per individual, and more accurate factor estimates positively impact the estimation of loadings and the regression coefficient matrix. However, as the variable dimensions $(p_1,p_2,p_3,p_4,p_5)$ gradually increase to $(250,250,250,100,50)$, there is a slight decrease in the estimation accuracy of the same modality variable-shared factor ($\V$) and study-shared loading ($\A$). This diminishing trend in $MT_{\V}$ and $MT_{\A}$ may be attributed to the increased complexity of estimation arising from the expanded dimensions, posing greater challenges. Finally, in Case 3.5 or 3.6 where the number of data sources or modalities are set to 20, i.e., $S=20$ (Case 3.5) or $M=20$ (Case 3.6), MMGFM remains dominant, while MSFR encounters breakdowns, and GFM and MRRR yield invalid estimates in Case 3.6. This underscores MMGFM's remarkable robustness and unparalleled adaptability to handle such intricate and complex scenarios.

\begin{table}[htbp]
  \centering
   \caption{Comparison of MMGFM  and other methods for parameter estimation.  Reported are the average (standard deviation) for performance metrics in Scenario  3, where nvec1=(50, 200), nvec2=(300, 200) and nvec3=(300, 400), pvec1=(250,50,50,100,50), pvec2=(250,50,250,100,50) and pvec3=(250,250,250,100,50).  Note that the algorithm of MSFR breaks down in Case 3.6, resulting in unavailable related results.}
 \resizebox{\textwidth}{!}{
    \begin{tabular}{llllllllllll}
     \hline
      &&Case 3.1&Case 3.2&\multicolumn{3}{c}{Case 3.3} &\multicolumn{3}{c}{Case 3.4 } &Case 3.5&Case 3.6 \\
      &&& &nvec1 &nvec2&nvec3 &pvec1 &pvec2&pvec3\\
      \cmidrule(lr){3-3} \cmidrule(lr){4-4}\cmidrule(lr){5-7}\cmidrule(lr){8-10}\cmidrule(lr){11-11}\cmidrule(lr){12-12}
 MMGFM &$MT_{\F}$
    &0.90  &0.90  &0.93  &0.96  &0.97  &0.83 &0.85 	&0.90  &0.86  &0.99  \\
          &       &(0.05)  &(0.06)  &(0.03)  &(0.01)  &(0.01)  &(0.06) &(0.05) &(0.05)
  &(0.04)  &(3E-3) \\
        &$MT_{\H}$ &0.99  &0.97  &0.98  &0.99  &0.99  &0.96 	&0.97&0.99
 &0.97  &1.00  \\
          &       &(5E-3) &(0.05)  &(0.01)  &(1E-3) &(7E-4) &(0.07) &(0.05) &(0.02)
  &(0.02)  &(2E-4) \\
          &$MT_{\V}$ &0.79  &0.86  &0.80  &0.80  &0.82  &0.76 &0.78 	&0.74
&0.79  &0.85  \\
          &       &(0.01)  &(0.01)  &(0.01)  &(0.01)  &(0.01)  &(0.01)  &(0.01) &(0.01) &(0.01)
 &(4E-3) \\
         &$MT_{\A}$ &0.88  &0.80  &0.87  &0.89  &0.90  &0.86 &0.87 	&0.85
&0.83  &0.94  \\
          &       &(0.03)  &(0.03)  &(0.02)  &(0.02)  &(0.02)  &(0.03)&(0.03)&(0.02)
&(0.03)  &(2E-3) \\
          &$MT_{\B}$ &0.74  &0.70  &0.75  &0.77  &0.79  &0.66 &0.70 	&0.72
&0.61  &0.88  \\
          &       &(0.04)  &(0.06)  &(0.03)  &(0.02)  &(0.02)  &(0.05)&(0.04) &(0.04)
&(0.03)  &(3E-3) \\
          &$ME_{\bb}$&0.48  &0.78  &0.67  &0.67  &0.67  &0.57&0.56 	&0.49
  &0.67  &0.53  \\
          &       &(0.01)  &(4E-3) &(0.01)  &(0.01)  &(4E-3)  &(0.01)&(0.01)&(0.01)
 &(0.01)  &(3E-3) \\
          \cmidrule(lr){3-12}
    GFM &$MT_{\F}$ &0.19  &0.09  &0.10  &0.11  &0.08  &0.10 	&0.10 &0.09  &0.19  &0.12  \\
          &       &(0.05)  &(0.07)  &(0.05)  &(0.08)  &(0.07)  &(0.03) &((0.07) &((0.03)
 &(0.06)  &(0.14)  \\
          &$MT_{\A}$ &0.29  &0.15  &0.18  &0.26  &0.26  &0.22 &0.21 &0.20
&0.24  &0.21  \\
          &       &(0.02)  &(0.05)  &(0.06)  &(0.05)  &(0.05)  &(0.05)&(0.04)&(0.04)
 &(0.07)  &(0.08)  \\
             \cmidrule(lr){3-12}
    MRRR&$MT_{\F}$ &0.55  &0.01  &0.08  &0.06  &0.06  &0.05 	&0.02 	&0.02
&0.11  &0.02  \\
          &       &(0.06)  &(4E-3) &(0.03)  &(0.01)  &(0.01)  &(0.02)&(0.01)&(0.01)
 &(0.01)  &(5E-3) \\
          & $MT_{\A}$&0.65  &0.02  &0.06  &0.06  &0.06  &0.07 &0.03 &0.03
&0.05  &0.04  \\
          &       &(0.05)  &(1E-3) &(0.01)  &(4E-3) &(2E-3) &(4E-3)	&(2E-3)	&(2E-3)
&(4E-3) &(1E-3) \\
          &$ME_{\bb}$ &0.23  &1.45  &1.51  &1.50  &1.50  &1.54 &1.49 	&1.51
 &1.50  &1.56  \\
          &       &(0.01)  &(0.03)  &(0.01)  &(0.01)  &(0.01)  &(0.01) &(0.03) &(0.03)
  &(0.01)  &(0.01)  \\
             \cmidrule(lr){3-12}
    MSFR& $MT_{\F}$&0.67  &0.71  &0.83  &0.86  &0.89  &0.70 	&0.73 	&0.71
&0.69   &- \\
          &       &(0.06)  &(0.09)  &(0.07)  &(0.06)  &(0.03)  &(0.04)&(0.06) &(0.07)
        &(0.04)  &(-)   \\
          &$MT_{\H}$ &0.57  &0.66  &0.81  &0.73  &0.77  &0.71&0.70&0.64
&0.75  &-  \\
          &      &(0.10)  &(0.09)  &(0.07)  &(0.08)  &(0.06)  &(0.07) &(0.07) &(0.09)
       &(0.02)  &(-)   \\
          &$MT_{\A}$ &0.80  &0.67  &0.85  &0.89  &0.91  &0.75 &0.75 	&0.74
&0.81  &-  \\
          &    &(0.08)  &(0.08)  &(0.05)  &(0.02)  &(0.01)  &(0.07)&(0.09) &(0.07)
&(0.06)  &(-)   \\
          &$MT_{\B}$ &0.53  &0.54  &0.64  &0.68  &0.75  &0.58&0.56 	&0.54
&0.48  &-  \\
          &   &(0.08)  &(0.08)  &(0.07)  &(0.08)  &(0.06)  &(0.07)&(0.07)&(0.08)
&(0.02)  &(-)   \\
          & $ME_{\bb}$ &0.47  &1.28  &0.73  &0.71  &0.70  &0.58 &0.73 	&0.59
&0.72  &-  \\
          &    &(0.01)  &(0.01)  &(0.01)  &(0.01)  &(5E-3) &(0.01)&(0.01) 	&(0.01)
&(0.01)  &(-)   \\
           \hline
    \end{tabular}%
    }
  \label{Scen_3}%
\end{table}%

\noindent\underline{\bf Scenario 4}. In this scenario, we evaluate the selection accuracy of the number of factors using the proposed step-wise SVR method. We investigate the performance under seven cases involving different sample sizes, levels of noise, variable dimensions, number of modalities or upper bounds on the number of factors. We follow the setttings of Case 3.3 in Scenario 3 with $\rho_m=3$, and consider seven cases. { Notably, Cases 4.6 and 4.7, which pertain to the insensitivity of the proposed step-wise SVR method, are relegated to Appendix D of the Supplementary Materials to conserve space; see Figure S1.}

\noindent\underline{\bf Case 4.1}. $(n_1, n_2, n_3)=(100, 150, 80),(p_1, p_2)=  (50, 150),p_3= 50,(p_4, p_5)= (100, 60)$;

\noindent\underline{\bf Case 4.2}. $(n_1, n_2, n_3)=(300, 200, 100),(p_1, p_2)=  (50, 150),p_3= 50,(p_4, p_5)= (100, 60)$;

\noindent\underline{\bf Case 4.3}. $(n_1, n_2, n_3)=(300, 200, 100),(p_1, p_2)=  (50, 150),p_3= 50,(p_4, p_5)= (100, 60)$ and $\varepsilon_{simj} \stackrel{i.i.d.}\sim N(0,3)$;

\noindent\underline{\bf Case 4.4}. $(n_1, n_2, n_3)=(300, 200, 100),(p_1, p_2)=  (250, 400),p_3= 250,(p_4, p_5)= (300,200)$;

\noindent\underline{\bf Case 4.5}. $(n_1, n_2, n_3)=(300, 200, 100)$, $p_m=150$ for $m\leq 7$, $p_m=50$ for $8<m\leq 13$ and $p_m=60$ for $13<m\leq 20$;

Note that $q=3$ and $\{q_{s}=2, s=1, \cdots, S\}$ for Scenarios 4.1-4.5. Figures \ref{fig:selectqSimu} and  S1 present the average and standard deviation of selected $q$ and $q_s$. The results illustrate that the step-wise SVR method can accurately identify the study-shared and study-specified  number of factors with a high correction rate, regardless of whether smaller or larger sample size, levels of noise, variable dimensions, number of modalities or upper bounds on the number of factors. We also observe that the selected correction rate exhibits an upward trend  as $n$ increases, whereas it declines  as $p$ or noise level  increases. These findings align with the  fact that  larger sample sizes provide more information for estimating model parameters, thereby improving the selection correction rate. Conversely,  the increased variable dimension or noise level heightens the complexity or weakens the signal strength of the model, thus reducing the selection accuracy.
\begin{figure}
  \centering
  {
  \includegraphics[width=1\textwidth]{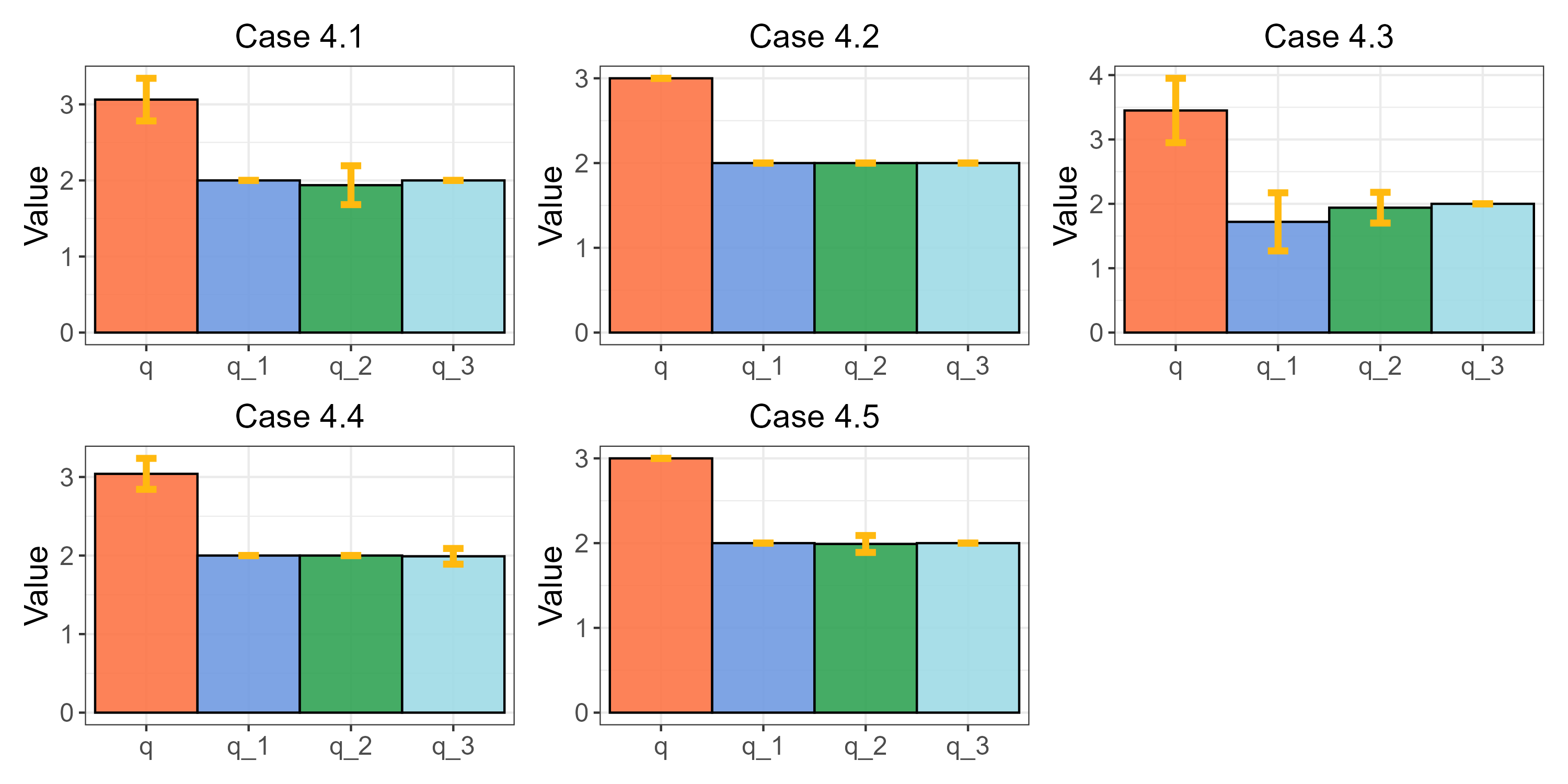}
  \label{fig:selectqSimu}
  }
  \caption{Bar plots of the average of selected number of factors for the proposed step-wise SVR method, based on 100 repetitions in Scenarios 4.1--4.5, with error bars representing the mean value $\pm$ standard deviation. Note that $q=3$ and $\{q_{s}=2, s=1, \cdots, 3\}$. }
\end{figure}
\nvs
{ To save space, Scenario 5 about the  computationally efficiency comparison and Scenarios 6 pertaining to the sensitivity analysis of MMGFM to the normality assumption of factors $\f_{si}$ and $\h_{si}$, are deferred to Appendix D of the Supplementary Materials; refer to Table S1 and Figures S2--S3 for the detailed results. }

\section{Real data application}\label{sec:real}
To demonstrate MMGFM's effectiveness on high-dimensional, multi-study, multi-modality datasets, {we analyze CITE-seq single-cell multimodal sequencing data from 12 PBMC subjects with varying COVID-19 status \citep{arunachalam2020systems}: four severe, three moderate, and five healthy. We divide the data into three groups (severe, moderate, healthy) as separate studies. After preprocessing (Appendix E in Supplementary Materials), we have 19318, 13258, and 30778 cells with 2666 genes and 39 protein markers per group. To control additional covariates, we regard age, sex and days since symptom onset as the augmented covariates. Gene expression data ($\X_{s1} \in \mathbb{R}^{n_s \times 2666}$) is in count form, while CLR-normalized protein marker data ($\X_{s2} \in \mathbb{R}^{n_s \times 39}$) is continuous~\citep{mule2022normalizing}.  Consequently, our analysis involves three distinct studies and two modalities.

Before applying MMGFM to this dataset, we require to determine the numbers of study-shared ($q$) and study-specific ($q_s$) factors. { We employ the criterion outlined in Section \ref{sec:modelselect} to obtain $\wh q=13$} and $(\wh q_1, \wh q_2, \wh q_3)=(3,3,3)$. To evaluate MMGFM's efficacy in feature extraction, we also implement its competitors (GFM, MRRR, MSFR, MultiCOAP) under the same factor numbers for impartial comparison.
Let $\F_s=(\f_{si}, \cdots, \f_{sn_s})^{\trans}, \H_s=(\h_{si}, \cdots, \h_{sn_s})^{\trans}$ and $\V_s=(v_{sim})\in \mathbb{R}^{n_s \times M}$. We define the features extracted by MMGFM as {$(\F_s, \H_s, \V_s)$}, 
whereas GFM and MRRR are limited to capturing only the study-shared factors. 
Additionally, MSFR and MultiCOAP are incapable of isolating the same modality variable-shared factor. 
Furthermore, MSFR is specifically designed for continuous modality data, while MultiCOAP  is limited to count-type modality. Accordingly, to ensure a methodologically appropriate comparison, we use $\X_{s2}$ and $\Z_s=(\z_{s1},\cdots,\z_{sn_s})^{\trans}$ as inputs for MSFR, while $\X_{s1}$ and $\Z_s$  are used for MultiCOAP,  aligning with each method's capabilities.

As we lack ground truth for factors and model parameters, we assess the association between the modality variables and the extracted features for each study. For the count-type modality, {we employ Poisson regression to regress each modality variable against the extracted features, calculating the MacFadden's R-square~\citep{mcfadden1972conditional, mcfadden1987regression} (R$^2_{s1j},j=1,2,\cdots, 2000$); for the continuous modality, we use linear regression
} and calculate the R-squares (R$^2_{s2j},j=1,2,\cdots, 39$). A higher R-squared value implies better performance. For stability, we  group the R-squares for each modality into ten groups and compute the average of each group.  The details of MacFadden's R-square and grouping are presented in Supplementary Materials.
Figure \ref{fig:R1}(a) shows MMGFM and MultiCOAP significantly outperform other approaches for count-type modalities, while MMGFM and MSFR excel for continuous modalities. Therefore, MMGFM comprehensively incorporates information from both modality types. To further show the rich information captured by MMGFM, we utilize the standard multi-omics analysis method~\citep{hao2021integrated} for each study, to identify  cell clusters, resulting 15, 14, and 15 cell clusters identified for three groups of subjects, respectively.  Figure \ref{fig:R1}(b) shows the adjusted MacFadden's R-square between these clusters and the extracted features, and suggests that both MMGFM and MultiCOAP demonstrate outstanding results in study 1, while  MMGFM significantly surpasses MultiCOAP in studies 2 and 3, highlighting the crucial significance of incorporating protein information into the analysis.

\begin{figure}
\centering
\includegraphics[width=0.9\textwidth ]{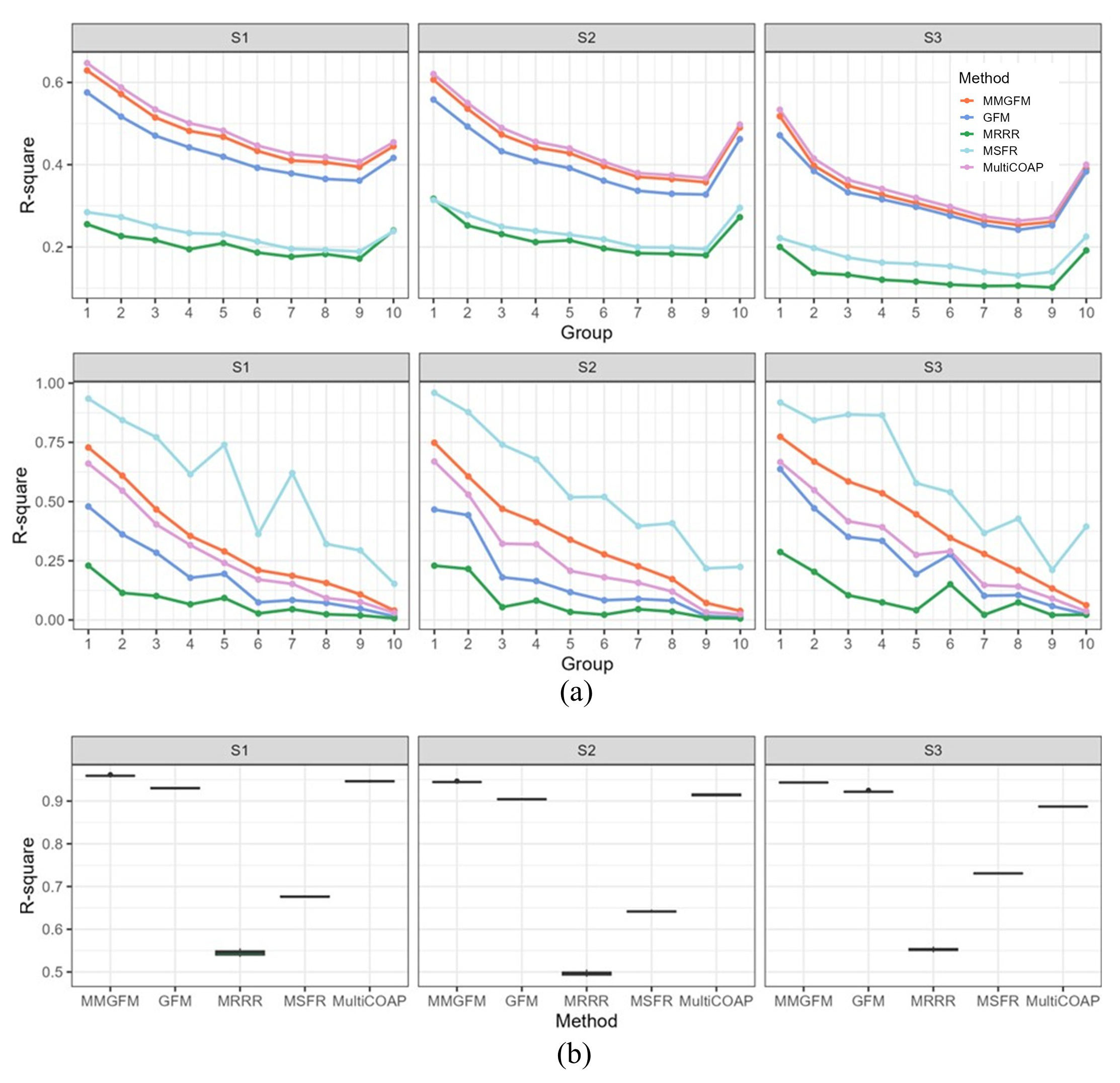}
  \caption{ Comparative analysis of MMGFM and other compared methods in feature extraction. (a) The R-squared values between the extracted features and the ten groups of variables in each modality (upper panel for gene modality, and bottom panel for protein modality) across three studies (S1--S3); (b) The adjusted MacFadden's R-squares between the extracted features and the annotated cell clusters across three studies. The computation is carried out ten times, with each time involving a random subsampling of 80\% of the cells from each study. }\label{fig:R1}
\end{figure}

MMGFM-extracted features enable joint clustering across three subjects and insights into biological mechanisms. We use Harmony \citep{korsunsky2019fast} to mitigate batch effects within these features. Leveraging batch-corrected MMGFM features, we perform integrative clustering with the Louvain algorithm \citep{stuart2019comprehensive}, identifying 16 shared cell clusters. Comparing these clusters with a standard tool confirms MMGFM's accuracy (Figure \ref{fig:R2}(a)\&(b)). We explore cluster variations by charting cluster frequency across three groups (Figure \ref{fig:R2}(c)). Clusters 1, 2, and 7 decrease in COVID-19 patients (S1-S2) vs. healthy (S3), while clusters 5, 8, and 9 increase from healthy to moderate, then severe cases. Differential expression analysis identifies marker genes/proteins (Figure \ref{fig:R2}(d)), matching them to a database (\url{http://xteam.xbio.top/CellMarker/}) to classify clusters: 1, 2, and 5 as T cells, monocytes, and B cells; 7-9 as monocyte-derived dendritic cells (mdDCs), macrophages, and plasmablasts. The proportions of T cell, monocyte, and mdDC  decrease in patients due to consumption, whereas the proportions of B cell, macrophage, and plasmablast increase attributed to immune response, consistent  with prior research \citep{arunachalam2020systems}.

Furthermore, we demonstrate that MMGFM's study-specific loading matrices aid in identifying differentially expressed genes/proteins across studies. By ranking magnitudes of gene-related ($\wh\B_{s1} \in \mathbb{R}^{2666\times 3}$) and protein-related ($\wh\B_{s2} \in \mathbb{R}^{39\times 3}$) loadings, we select top genes/proteins for each of the three directions ($q_s=3$) and visualize their expression (Figure \ref{fig:R2}(e), Supplementary Figure S3). This methodology is driven by the observation that the magnitude of the loading $\hat b_{smjk}$ indicate the contribution of gene/protein $j$ to the $k$-th study-specific factor.  Figure \ref{fig:R2}(e) highlights genes/proteins specific to severe COVID-19, like {\it IGKC, IGHG1, IGLC2} (in antibody production) and {\it CD3, CD4, CD28} (in T-cell activation), with higher expression in severe vs. healthy/moderate cases, reflecting immune hyperactivation. The study-specific loading matrices also enhance gene/protein co-expression network analysis, as shown in Appendix E.2 of Supplementary Materials.
}

\begin{figure}
\centering
\includegraphics[width=0.8\textwidth]{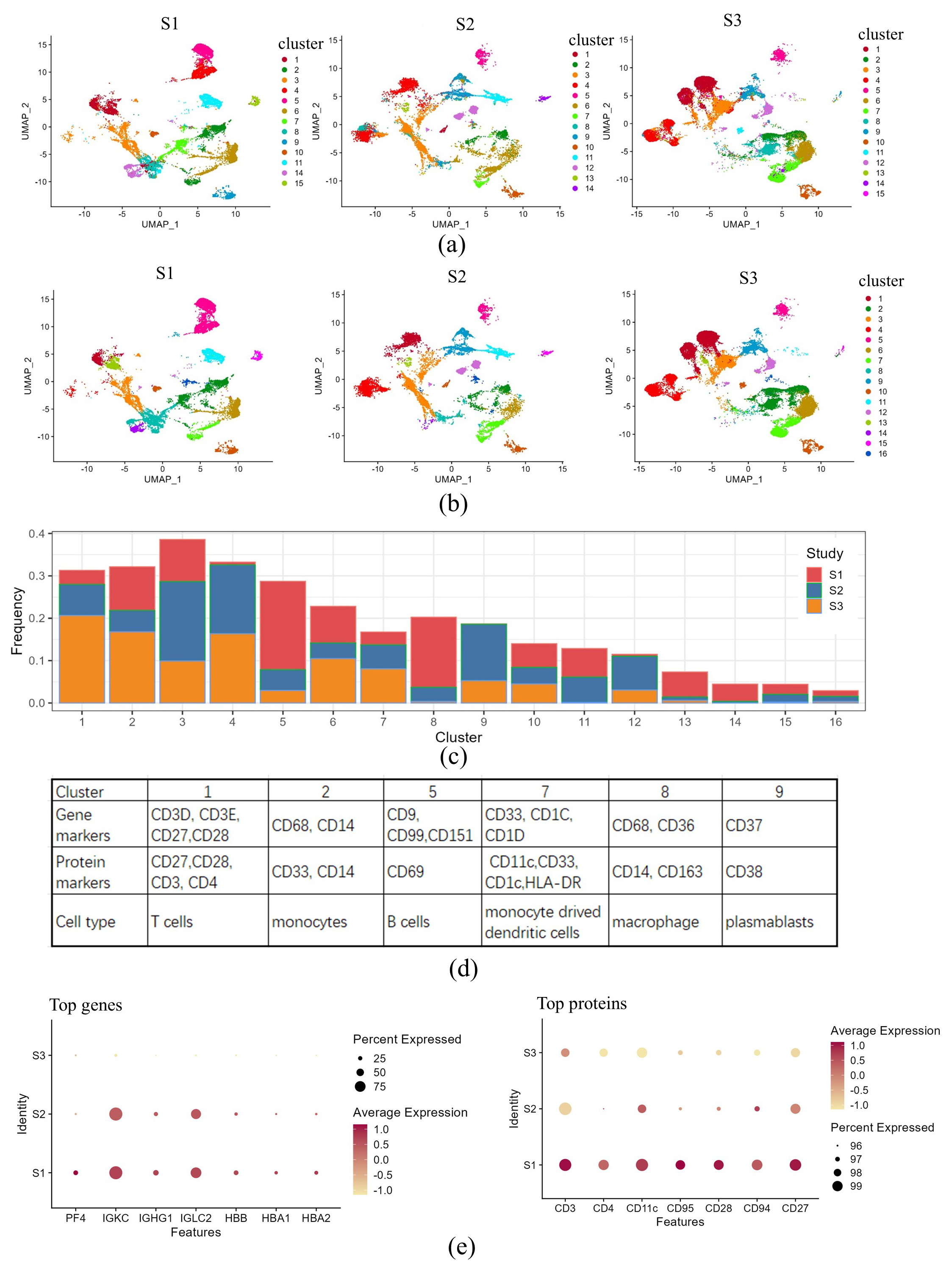}
  \caption{Downstream applications utilizing the features extracted by MMGFM. { (a) UMAP plot for visualizing the annotated cell clusters identified by the standard analysis tool, where the UMAP projections are calculated based on the batch-corrected features extracted by MMGFM. (b) UMAP plot for visualizing the cell clusters that are identified using MMGFM's batch-corrected features. Note the
methods used to obtain the cell clusters differ between (a) and (b);} (c) Barplot of frequencies of cell clusters identified based on MMGFM's batch-corrected features across three studies. (d) The gene/protein markers and annotated cell types for clusters 1, 2, 5 and 7-9. { (e) Dot plots of top genes and proteins based on the magnitudes of the loading matrices specific to study 1 (S1: COVID-19 patients with severe symptom)}. }\label{fig:R2}
\end{figure}

\nvs
\section{Discussion}\label{sec:dis}
We introduce MMGFM, a novel covariate-augmented generalized factor model, to integrate multi-study, multi-modality data, addressing a gap in existing models that focus on either multi-study or multi-modality integration. We ensure model identifiability and use variational approximation to handle high-dimensional nonlinear integration, developing a computationally efficient variational EM algorithm with linear complexity. By profiling variational parameters, we establish estimator consistency and convergence rates. A step-wise singular value ratio-based method determines optimal factor numbers. Numerical experiments highlight MMGFM's significant advantage over existing methods.

There are several intriguing research avenues that warrant further consideration. Firstly, we are currently unable to establish asymptotical results in the more arduous scenario where $p$ is comparable to or even exceeds $\min_s n_s$. This limitation arises from our framework's treatment of factors as randomly distributed, rather than fixed but unknown parameters, which renders the technical tool of profile M-estimation inapplicable in a straightforward manner. Secondly, the consistency of factor number selection remains an unresolved challenge.
{ Lastly, we have not addressed the removal of batch effects in the feature space among studies, which poses an intriguing question for future research.}
\nvs

\section*{Funding}
Liu's work was supported by National Natural Science Foundation of China  (12401361), Natural Science Foundation of Sichuan Province (2025ZNSFSC0809) and  the Fundamental Research Funds for the Central Universities (1082204112J06). Zhong's work was supported by National Natural Science Foundation of China  (12401349) and  China Postdoctoral Science Foundations (2024M751116).
\newpage
\bibliographystyle{apalike}

\bibliography{reflib}

\begin{thebibliography}{}

\bibitem[Argelaguet et~al., 2018]{argelaguet2018multi}
Argelaguet, R., Velten, B., Arnol, D., Dietrich, S., Zenz, T., Marioni, J.~C.,
  Buettner, F., Huber, W., and Stegle, O. (2018).
\newblock Multi-omics factor analysis—a framework for unsupervised
  integration of multi-omics data sets.
\newblock {\em Molecular systems biology}, 14(6):e8124.

\bibitem[Arunachalam et~al., 2020]{arunachalam2020systems}
Arunachalam, P.~S., Wimmers, F., Mok, C. K.~P., Perera, R.~A., Scott, M.,
  Hagan, T., Sigal, N., Feng, Y., Bristow, L., Tak-Yin~Tsang, O., et~al.
  (2020).
\newblock Systems biological assessment of immunity to mild versus severe
  covid-19 infection in humans.
\newblock {\em Science}, 369(6508):1210--1220.

\bibitem[Avalos-Pacheco et~al., 2022]{avalos2022heterogeneous}
Avalos-Pacheco, A., Rossell, D., and Savage, R.~S. (2022).
\newblock Heterogeneous large datasets integration using bayesian factor
  regression.
\newblock {\em Bayesian Analysis}, 17(1):33--66.

\bibitem[Chandra et~al., 2024]{chandra2024inferring}
Chandra, N.~K., Dunson, D.~B., and Xu, J. (2024).
\newblock Inferring covariance structure from multiple data sources via
  subspace factor analysis.
\newblock {\em Journal of the American Statistical Association},
  (just-accepted):1--25.

\bibitem[De~Vito and Avalos-Pacheco, 2023]{de2023multi}
De~Vito, R. and Avalos-Pacheco, A. (2023).
\newblock Multi-study factor regression model: an application in nutritional
  epidemiology.
\newblock {\em arXiv preprint arXiv:2304.13077}.

\bibitem[De~Vito et~al., 2019]{de2019multi}
De~Vito, R., Bellio, R., Trippa, L., and Parmigiani, G. (2019).
\newblock Multi-study factor analysis.
\newblock {\em Biometrics}, 75(1):337--346.

\bibitem[Hao et~al., 2021]{hao2021integrated}
Hao, Y., Hao, S., Andersen-Nissen, E., Mauck, W.~M., Zheng, S., Butler, A.,
  Lee, M.~J., Wilk, A.~J., Darby, C., Zager, M., et~al. (2021).
\newblock Integrated analysis of multimodal single-cell data.
\newblock {\em Cell}, 184(13):3573--3587.

\bibitem[Korsunsky et~al., 2019]{korsunsky2019fast}
Korsunsky, I., Millard, N., Fan, J., Slowikowski, K., Zhang, F., Wei, K.,
  Baglaenko, Y., Brenner, M., Loh, P.-r., and Raychaudhuri, S. (2019).
\newblock Fast, sensitive and accurate integration of single-cell data with
  harmony.
\newblock {\em Nature methods}, 16(12):1289--1296.

\bibitem[Li and Li, 2022]{li2022integrative}
Li, Q. and Li, L. (2022).
\newblock Integrative factor regression and its inference for multimodal data
  analysis.
\newblock {\em Journal of the American Statistical Association},
  117(540):2207--2221.

\bibitem[Liu et~al., 2023a]{liu2023probabilistic}
Liu, W., Liao, X., Luo, Z., Yang, Y., Lau, M.~C., Jiao, Y., and et~al. (2023a).
\newblock Probabilistic embedding, clustering, and alignment for integrating
  spatial transcriptomics data with precast.
\newblock {\em Nature Communications}, 14(1):296.

\bibitem[Liu et~al., 2023b]{GFMLiu}
Liu, W., Lin, H., Zheng, S., and Liu, J. (2023b).
\newblock Generalized factor model for ultra-high dimensional correlated
  variables with mixed types.
\newblock {\em Journal of the American Statistical Association},
  118(542):1385--1401.

\bibitem[Liu and Zhong, 2024a]{liu2024highdimensional}
Liu, W. and Zhong, Q. (2024a).
\newblock High-dimensional covariate-augmented overdispersed multi-study
  poisson factor model.
\newblock {\em arXiv preprint arXiv:2408.10542}.

\bibitem[Liu and Zhong, 2024b]{liu2024high}
Liu, W. and Zhong, Q. (2024b).
\newblock High-dimensional covariate-augmented overdispersed poisson factor
  model.
\newblock {\em Biometrics}, 80(2):ujae031.

\bibitem[Luo et~al., 2018]{luo2018leveraging}
Luo, C., Liang, J., Li, G., Wang, F., Zhang, C., Dey, D.~K., and Chen, K.
  (2018).
\newblock Leveraging mixed and incomplete outcomes via reduced-rank modeling.
\newblock {\em Journal of Multivariate Analysis}, 167:378--394.

\bibitem[McFadden, 1972]{mcfadden1972conditional}
McFadden, D. (1972).
\newblock Conditional logit analysis of qualitative choice behavior.

\bibitem[McFadden, 1987]{mcfadden1987regression}
McFadden, D. (1987).
\newblock Regression-based specification tests for the multinomial logit model.
\newblock {\em Journal of econometrics}, 34(1-2):63--82.

\bibitem[Mul{\`e} et~al., 2022]{mule2022normalizing}
Mul{\`e}, M.~P., Martins, A.~J., and Tsang, J.~S. (2022).
\newblock Normalizing and denoising protein expression data from droplet-based
  single cell profiling.
\newblock {\em Nature communications}, 13(1):2099.

\bibitem[Pang et~al., 2023]{pang2023factor}
Pang, D., Zhao, H., and Wang, T. (2023).
\newblock Factor augmented inverse regression and its application to microbiome
  data analysis.
\newblock {\em Journal of the American Statistical Association}, pages 1--11.

\bibitem[Stenlund et~al., 2018]{stenlund2018successful}
Stenlund, T., Lyr{\'e}n, P.-E., and Ekl{\"o}f, H. (2018).
\newblock The successful test taker: exploring test-taking behavior profiles
  through cluster analysis.
\newblock {\em European Journal of Psychology of Education}, 33:403--417.

\bibitem[Stuart et~al., 2019]{stuart2019comprehensive}
Stuart, T., Butler, A., Hoffman, P., Hafemeister, C., Papalexi, E., Mauck,
  W.~M., Hao, Y., Stoeckius, M., Smibert, P., and Satija, R. (2019).
\newblock Comprehensive integration of single-cell data.
\newblock {\em Cell}, 177(7):1888--1902.

\bibitem[Vandereyken et~al., 2023]{vandereyken2023methods}
Vandereyken, K., Sifrim, A., Thienpont, B., and Voet, T. (2023).
\newblock Methods and applications for single-cell and spatial multi-omics.
\newblock {\em Nature Reviews Genetics}, pages 1--22.

\bibitem[Wang and Blei, 2019]{wang2019frequentist}
Wang, Y. and Blei, D.~M. (2019).
\newblock Frequentist consistency of variational bayes.
\newblock {\em Journal of the American Statistical Association},
  114(527):1147--1161.

\bibitem[Welch et~al., 2019]{welch2019single}
Welch, J.~D., Kozareva, V., Ferreira, A., Vanderburg, C., Martin, C., and
  Macosko, E.~Z. (2019).
\newblock Single-cell multi-omic integration compares and contrasts features of
  brain cell identity.
\newblock {\em Cell}, 177(7):1873--1887.

\end{thebibliography}

\end{document}